\documentclass[preprint,12pt]{elsarticle}

\usepackage{amssymb}

\usepackage{amsmath}
\usepackage{bm}
\usepackage{comment}
\usepackage{amssymb}
\usepackage{empheq}

\newcommand{\Ev}[2]{{\mathbb E}_{{#1}} \left[ #2 \right]}
\newcommand{\corrto}{ \mathrel{\widehat{=}} }

\journal{Physica A}

\begin{document}

\begin{frontmatter}



\title{An eikonal equation approach to thermodynamics and
the gradient flows in information geometry}

\author[1st]{Tatsuaki Wada}
\author[2nd]{Antonio M. Scarfone}
\author[last]{Hiroshi Matsuzoe}

\address[1st]{
Region of Electrical and Electronic Systems Engineering, Ibaraki University, Nakanarusawa-cho, Hitachi 316-8511, Japan}
\address[2nd]{Istituto dei Sistemi Complessi, Consiglio Nazionale delle Ricerche (ISC-CNR), c/o Politecnico di Torino, 10129 Torino, Italy}
\address[last]{
Department of Computer Science and Engineering, Nagoya Institute of Technology, Gokiso-cho, Showa-ku, Nagoya, 466-8555, Japan
}

\begin{abstract}
We can incorporate a "time" evolution into thermodynamics as a Hamilton-Jacobi dynamics.
A set of the equations of states in thermodynamics is considered as the generalized eikonal equation,
which is equivalent to Hamilton-Jacobi equation.
We relate the Hamilton-Jacobi dynamics of a simple thermodynamic system to 
the gradient flows in information geometry.
\end{abstract}


\end{frontmatter}


\section{Introduction}
\label{intro}

Since classical thermodynamics studies equilibrium states and transitions between them, it is a common belief that there is \textit{no dynamics} in thermodynamics, unlike
Lagrange or Hamilton dynamics in classical mechanics.
Nevertheless there is an analogy between classical mechanics and thermodynamics.
For example, Rajeev \cite{Rajeev} formulated classical thermodynamics in terms of Hamilton-Jacobi theory,
Vaz \cite{Vaz} studied a Lagrangian description of thermodynamics, and Baldiotti et al. \cite{BFM16} showed a Hamiltonian approach
to thermodynamics.

Similarly to the case of thermodynamics, there is no concept of "time" in the information geometry (IG) \cite{Amari}.
During our studies \cite{WMS15} on the information geometric structures in generalized thermostatics, we are inspired by a remarkable work \cite{Giovanni} by Prof. Pistone. He has derived the Euler Lagrange equations from the Lagrange action integral by computing the "velocity" and "acceleration" of a one-dimensional statistical model. 
The quite remarkable points of his results are as follows.
\begin{itemize}
\item[i)] introducing the concepts of ``velocity'' as Fisher's score, and ``acceleration'' as a second order geometry, so that
  a flat structure in IG is characterized as \textit{an orbit along the curve with zero-acceleration}.
  \item[ii)]
    introducing the dynamics of a density $q(t)$
    in IG as a gradient flow equation \cite{Giovanni15}, 
\begin{align}
  \frac{d}{dt} \ln \frac{q(t)}{q_2} = -\nabla \Big( q \to \mathrm{D}(q(t) \Vert q_2) \Big) =
- \ln \frac{q(t)}{q_2} + \mathrm{D}(q(t) \Vert q_2),
  \label{GradflowEq}
\end{align}
where  $\mathrm{D}(q(t) \Vert q_2)$ is the Kullback-Leibler divergence
\begin{align}
 \mathrm{D}(q(t) \Vert q_2) := \Ev{q(t)}{ \ln \frac{q(t)}{q_2} }.
\end{align}
The solution of Eq. \eqref{GradflowEq} was given by
\begin{align}
  q(t) = \exp \Big( \mathrm{e}^{-t} \ln q_0 + (1-\mathrm{e}^{-t}) \ln q_2 - \Psi(t) \Big),
  \label{q(t)}
\end{align}
where $\Psi(t)$ is the normalization function.
This solution $q(t)$ evolves smoothly from $ q_0 := q(0)$ to $q_2:= q(\infty)$ with increasing the parameter $t$.
For more technical details see for example \cite{Pistone13,MP15,Giovanni15}.
\end{itemize}
As is well-known, the dynamics of a point particle is described by Newton's mechanics, in which the zero-acceleration of a particle means that there is no external force acting on the particle.  
Point i) reminds us Einstein's general relativity \cite{Dirac}. It states that
a particle which is free from an external non-gravitational force
always moves along a geodesic in a curved space-time. 
Indeed, for a curve $\gamma(t)$ on a smooth manifold $\mathcal{M}$ with an affine connection $\nabla$, the geodesic equation is
described by
\begin{align}
  \nabla_{\frac{d \gamma}{dt}} \frac{d \gamma(t)}{d t} = 0,
\end{align}
which states that there is no tangent component of the acceleration vector of the geodesic curve $\gamma(t)$.
This consideration leads us to study a relation among analytical mechanics, thermodynamics and IG. 

On Point ii), we wonder the functional form of the parameter $t$ dependency, i.e., the double exponential $t$-dependency of $\exp( \exp(-t))$ in \eqref{q(t)}. Since the standard formalism of IG has no concept of time-evolution,
a natural question
is that where does this  double exponential $t$-dependency come from? In other words, what is the physical meaning of the ``time like'' parameter $t$ in the gradient-flow equation in IG?
This is the main motivation of this contribution.

The gradient flows in IG were studied by
Fujiwara \cite{F93}, Fujiwara and Amari \cite{FA95}, and Nakamura \cite{N94}, in which they found that the gradient systems on manifolds of even dimensions are completely integrable Hamiltonian systems. This relation between the gradient systems and Hamilton systems is somewhat mysterious and incomprehensible.
Later, Boumuki and Noda \cite{BN16} gave a theoretical explanation for this  relationship between Hamiltonian flow and gradient flow from the viewpoint of symplectic geometries.
In this contribution we consider the differential equations:
\begin{align}
  \frac{d \theta^i(t)}{dt} &= - \theta^i(t), 
  \label{grad-eqs}
\end{align}
which is equivalent to the gradient flow equations
\begin{align}
  \frac{d \eta_i(t)}{dt} &= - g_{ij}(\bm{\theta}) \, \frac{\partial }{\partial \eta_j(t)} \Psi^{\star}(\bm{\eta}),
\end{align}
for the potential functions $\Psi(\bm{\theta})$ and $\Psi^{\star}(\bm{\eta})$  in a dually-flat statistical manifold $(\mathcal{M}, g, \nabla, \nabla^{\star})$.
We here take a different approach to the gradient flows in IG. 
Since the relation between the Legendre transformations in a simple thermodynamical system and those in IG are already known (e.g, section 2 in \cite{WMS15}),  our discussion is based on this correspondence between thermodynamic macroscopic variables and affine coordinates in IG.
Basically in both (equilibrium) thermodynamics and IG, there is no time parameter.
Nevertheless we can introduce a "time" parameter $\tau$ 
\footnote{In order to avoid possible confusion, we use $\tau$ as a "time" (or mock time) parameter to distinguish it from the parameter $t$ which is used in gradient equations. 
}
as the parameter which describes Hamilton-Jacobi (HJ) dynamics.
By generalizing the eikonal equation in optics, which is equivalent to the square of HJ equation,
we can describe a "time" ($\tau$) evolution of thermodynamic systems. 

The rest of the paper consists as follows.
The next section briefly reviews the basic of geometrical optics. The eikonal equation describes a real optical path between two different points in an optical media. The generalized eikonal equation is introduced and the relations with HJ equation are explained. In section 3, a "time" parameter $\tau$ is introduced in the standard settings of equilibrium thermodynamics. Einstein's \textit{vielbein} formalism (or Cartan's \textit{method of moving frame}) plays a central role in order to obtain a Riemann metric in an equilibrium thermodynamic system. By studying the gradient flow in the ideal gas model, we will relate the parameter $t$ with the temperature $T$ of the thermodynamic system. As a more realistic gas model, the gas model by van der Waals is also studied.
Section 4 discusses two main issues.
In the first subsection 4.1 we discuss a simple canonical probability of a thermal system, and show the relation to the gradient flow equation \eqref{GradflowEq}.
The origin of the double exponential $t$-dependency of the solution \eqref{q(t)} is explained as the $\beta$-dependency of the canonical probability.
Relation to Gompertz function is also shown.
In the next subsection 4.2, we consider a very simple model described by the specific characteristic function $W$, and
discuss the relation between the gradient flow and Hamilton flow.
Final section is devoted to our conclusion. In Appendix A, HJ equation is derived by applying appropriate canonical transformation. Appendix B explains a constant (effective) pressure process in the ideal and van der Waals gas models.
We use Einstein summation convention throughout this paper except in Subsection \ref{subsec}.


\section{Geometrical optics and generalized eikonal equation}
\label{sec:pre}
In geometrical optics, the real path of the ray in a media with refractive index $n(\bm{q})$ is characterized by Hamilton's characteristic function, (or point eikonal function),
\begin{align}
  I(\bm{q}; \bm{q}_0) &= \int_{s_0}^s n(\bm{q}(s)) \left\vert \frac{d \bm{q}(s)}{d s} \right\vert ds
=: \int_{s_0}^s L_{\rm op} \left( \bm{q}(s), \frac{d \bm{q}(s)}{d s} \right) ds,
\end{align}
where $s$ is a curve (path) parameter and $L_{\rm op}$ is the so called optical Lagrangian. 
The point eikonal function $I(\bm{q}; \bm{q}_0)$ is the length of a real optical path from a point $\bm{q}_0 := \bm{q}(0) = (x_0, y_0, z_0) $ to another point $\bm{q}(s) = (x,y,z)$.
From Fermat's principle (the principle of least time), the real optical path satisfies $\delta I =0$. Then the variation $\delta I$ of $I(\bm{q}; \bm{q}_0)$ with respect to the infinitesimal variations of the two end points $\bm{q}$ and $\bm{q}_0$ is given by
\begin{align}
   \delta I =  \left[ \frac{\partial L_{\rm op}}{ \partial \left( \frac{d  \bm{q}}{d s} \right)} \cdot \delta \bm{q}\right]_{s_0}^s 
= \Big[ \bm{p} \cdot \delta \bm{q} \Big]^s_{s_0} = \bm{p}(\bm{q}) \cdot \delta\bm{q} - \bm{p}(\bm{q}_0) \cdot \delta\bm{q}_0\; . 
\end{align}
Consequently if we find the point eikonal function $I(\bm{q}; \bm{q}_0)$, the ray vector $\bm{p}(\bm{q})$ at a position $\bm{q}$ and that $\bm{p}(\bm{q}_0)$ at the other point $\bm{q}_0$ are obtained by
\begin{align}
  \bm{p}(\bm{q}) = \frac{\partial I}{\partial \bm{q}}, \quad
  \bm{p}(\bm{q}_0) = -\frac{ \partial I}{\partial \bm{q}_0},
\end{align}
respectively. Since $\vert \bm{p}(\bm{q}) \vert^2 = n^2(\bm{q}) $, the point eikonal function $I(\bm{q}; \bm{q}_0)$ satisfies the
eikonal equations
\begin{subequations}
\begin{align}
    \left( \frac{\partial I}{\partial x} \right)^2 + \left( \frac{\partial I}{\partial y} \right)^2 + \left( \frac{\partial I}{\partial z} \right)^2 &= n^2(\bm{q}),\\
 \left( \frac{\partial I}{\partial x_0} \right)^2 + \left( \frac{\partial I}{\partial y_0} \right)^2 + \left( \frac{\partial I}{\partial z_0} \right)^2 &= n^2(\bm{q}_0).
\end{align}
\label{Eikonal}
\end{subequations}
In order to determine the real optimal path between $\bm{q}$ and $\bm{q}_0$ the both relations \eqref{Eikonal} are needed. 
Recall that the eikonal equation is obtained from the wave equation in the limit of short wavelength $\lambda$, i.e.,
$\vert \nabla n \vert \ll k = 2 \pi / \lambda$. Historically Hamilton invented his formalism of classical mechanics from his theory of systems of rays \cite{Hamilton}.

Now let us consider the following generalized eikonal equation 
\begin{align}
    g^{\mu \nu}(\bm{q}) \left( \frac{\partial W}{\partial q^{\mu}} \right) \left( \frac{\partial W}{\partial q^{\nu}} \right) = E^2(\bm{q}), \quad \mu, \nu=1,2,\ldots,N,
\label{GEikonal}
\end{align}
in a  smooth manifold with $N$-dimension. 
Here $W=W(\bm{q}, \bm{P})$ is Hamilton's characteristic function,  $g^{\mu \nu}(\bm{q})$ is the inverse tensor of a metric $g_{\mu \nu}(\bm{q})$ of the manifold, and $E(\bm{q})$ takes a real positive value.  It is known that $W(\bm{q}, \bm{P})$ is a generating function which relates the original variables $(\bm{q}, \bm{p})$ and new variables $(\bm{Q}, \bm{P})$ by the relations 
\begin{align}
  p_{\mu} = \frac{\partial W}{\partial q^{\mu}}, \quad Q^{\mu} = \frac{\partial W}{\partial P_{\mu}},
 \quad \mu=1,2, \cdots, N.
  \label{p_Q}
\end{align}
Note that if $g^{\mu \nu}(\bm{q})=\delta^{\mu \nu}$, then the generalized eikonal equation \eqref{GEikonal} reduces to the standard eikonal equation \eqref{Eikonal} with $E = n(\bm{q}) = \vert \bm{p}(\bm{q}) \vert$.
We see that taking the square root of the generalized eikonal equation \eqref{GEikonal} leads to
the Hamilton-Jacobi (HJ) equation \cite{HJeq}
\begin{align}
  H\left( \bm{q},  \frac{\partial W}{\partial \bm{q}} \right)  - E(\bm{q}) = 0,
  \label{HJeq}
\end{align}
with the time-independent Hamiltonian
\begin{align}
   H( \bm{q}, \bm{p}) = \sqrt{ g^{\mu \nu}(\bm{q}) p_{\mu} p_{\nu}} \; .
  \label{H}
\end{align} 
It is also known that for any time-independent Hamiltonian, by separation of variables,
the corresponding action $S$ is expressed as
\begin{align}
 S( \bm{q}, \bm{P}, \tau) = W( \bm{q}, \bm{P}) - E(\bm{P}) \, \tau,
 \label{action}
\end{align}
where $\tau$ is a "time" parameter, $E(\bm{P})$ is a total energy of the Hamiltonian $H$ as a function of $\bm{P}$.
In addition, the action $S(\bm{q}, \bm{P}, \tau)$
is a generating function of the (type-2) canonical transformation from the original variables $(\bm{q}, \bm{p})$ to new variables $(\bm{Q}, \bm{P})$ which are conserved, i.e.,
\begin{align}
  \frac{d Q^{\mu}}{d\tau} = 0, \quad \frac{d P_{\mu}}{d\tau} = 0, \quad \mu=1,2, \cdots, N.
  \label{Q_P}
\end{align}
The transformed new Hamiltonian $K$ is given by
\begin{align}
  K(\bm{Q}, \bm{P}) = H\left( \bm{q},  \frac{\partial W}{\partial \bm{q}} \right)  - E = 0,
  \label{K}
\end{align}
which is equivalent to HJ equation \eqref{HJeq}, and the relations \eqref{p_Q}
are satisfied (see \ref{HJbyCT}).

It is worth noting that 
\begin{align}
      p_{\mu} \, \frac{d q^{\mu}}{d \tau}  = p_{\mu}\, \frac{\partial H}{\partial p_{\mu}} = \frac{g^{\mu \nu}(\bm{q}) p_{\mu} p_{\nu}}{\sqrt{ g^{\rho \lambda}(\bm{q}) p_{\rho} p_{\lambda}}} = H,
\end{align}
where in the first step we used $d q^{\mu}/ d\tau = \partial H / \partial p_{\mu}$.
Consequently the corresponding Lagrangian $\mathcal{L}$, which is the Legendre dual of the Hamiltonian \eqref{H}, becomes null,
\begin{align}
     \mathcal{L} \left(\bm{q}, \frac{d \bm{q}}{d \tau} \right) := p_{\mu}\, \frac{d q^{\mu}}{d \tau} -H( \bm{q}, \bm{p}) = 0.
\end{align}
This leads to
\begin{align}
  dS(\bm{q}, \bm{P}, \tau)  = \mathcal{L} \left(\bm{q}, \frac{d \bm{q}}{d \tau} \right) \, d\tau = 0,
\end{align}
and
by using Eq. \eqref{action} we see that
\begin{align}
  dW(\bm{q}, \bm{P})  = E(\bm{P})  d\tau.
 \label{dW}
\end{align}

In addition,
according to Carath\'{e}odry \cite{Snow67}, we introduce the generating function
\begin{align}
   G(\bm{q}, \tau; \bm{q}_0, \tau_0) := S(\bm{q}, \bm{P}, \tau)-S(\bm{q}_0, \bm{P}, \tau_0),
  \label{G}
\end{align}
which generates the canonical transformation between a set of canonical coordinates  $(\bm{q}_0 :=\bm{q}(\tau_0), \bm{p}_0 :=\bm{p}(\tau_0))$ and another set of canonical coordinates $(\bm{q}(\tau), \bm{p}(\tau))$ via the conserved quantities $(\bm{Q}, \bm{P})$.
Note that $S(\bm{q}_0, \bm{P}, \tau_0)$ is the generating function of the canonical transformation between $(\bm{q}_0, \bm{p}_0)$ and $(\bm{Q}, \bm{P})$ and
$-S(\bm{q}, \bm{P}, \tau)$ is that of the canonical transformation between  $(\bm{Q}, \bm{P})$ and
$(\bm{q}, \bm{p})$.
Later we shall use this generating function $G$ in order to obtain the "time" dependency or dynamics described by the $\tau$-parameter.

\section{Thermodynamical systems}
\label{sec:TDsys}
In general, equilibrium thermodynamic systems with $N$-independent macroscopic variables are completely described by the $N$-independent equations of states, which can be cast into the following form  \cite{Vaz},
\begin{align}
  e_i{}^{\mu} ({\bm q}) \, p_{\mu} = r_i, \quad i, \mu =1,2, \ldots N,
  \label{vielbein}
\end{align}
where each $p_{\mu}$ is an intensive thermodynamic variable, $q_{\mu}$ is extensive one, and 
$r_i$ are independent constants. We assume that the matrix $e_i{}^{\mu} ({\bm q})$ to be invertible so that 
we have
\begin{align}
  p_{\mu} =  e_{\mu}{}^i  ({\bm q})  \; r_i,
 \label{p}
\end{align}
where $e_{\mu}{}^i(\bm q)$ is the inverse matrix of $e_i{}^{\mu} ({\bm q})$ and the following relations are satisfied.
\begin{align}
  e_{\mu}{}^i  ({\bm q})  \; e_j{}^{\mu} ({\bm q}) = \delta^i_j, \quad
  e_{\nu}{}^i   ({\bm q}) \; e_i{}^{\mu} ({\bm q}) = \delta^{\mu}_{\nu},
\end{align}
where $\delta^i_j$ and $\delta^{\mu}_{\nu}$ denote Kronecker's delta.
Since each $r_i$ is independent, we can assign $\{ r_i \}$ as the components in an orthogonal system with the invertible constant matrix $\eta^{i j}$, and $ e_i{}^{\mu} $ can be thought of as a \textit{vielbein} \cite{Einstein}. In other words,  the frame consisting of the orthogonal basis $\{ r_i \}$ is Cartan's moving frame.
The vielbein (or N-bein) \cite{Einstein} is used in the field of general relativity \cite{Brau}.
In the fields of IG, to the best of our knowledge, the first trial of applying vielbein formalism is Ref. \cite{W19}.
In general the frame of $\{p_{\mu} \}$ on a manifold are non-orthogonal and characterized by a metric tensor $g$.
The vielbein field $e_i{}^{\mu} ({\bm q})$ relates this non-orthogonal frame of $\{p_{\mu} \}$ with the local orthogonal frame of $\{ r_i \}$. In order to distinguish the two different frames, Greek and Latin indices are often used, and we follow this convention here.

The inner product in the orthogonal frame of
$\{ r_i \}$ with a diagonal metric tensor $\eta$
and that in the frame of $\{ p_{\mu} \}$ with a metric tensor $g$ are related with
\begin{align}
   g^{\mu \nu} \, p_{\mu} p_{\nu} = \eta^{ij} \, r_i r_j.
   \label{inner}
\end{align}
Here we regard this relation as the generalized eikonal equation \eqref{GEikonal} with the refractive index $n(\bm{q}) =  \eta^{ij} r_i r_j$,
which is a constant and consequently this corresponds to a homogeneous isotropic medium.

From Eq. \eqref{vielbein} and Eq. \eqref{inner}, it follows the next relation
\begin{align}
   g^{\mu \nu} = \eta^{i j}  e_i{}^{\mu}  e_j{}^{\nu},
  \label{R-metric}
\end{align}
where $g^{\mu \nu}$ is the inverse matrix of the metric $g_{\mu \nu}$ on the $N$-dimensional smooth manifold $\mathcal{M}$, i.e., they satisfy 
\begin{align}
  g^{\mu \rho}  \, g_{\rho \nu} = \delta^{\mu}_{\nu}.
\end{align}
It is worth noting that the matrix elements $e_i{}^{\mu} $ of a vielbein is determined by $N^2$ components, whereas those $g_{\mu \nu}$ of a Riemann metric  have $N(N+1)/2$ components.
Consequently, the metric $g_{\mu \mu}$ is obtained from a given vielbein as
\begin{align}
   g_{\mu \nu} = \eta_{i j} \,  e_{\mu}{}^{i}  e_{\nu}{}^{j},
\end{align}
however, the converse is not possible in general.

In this way, by using vielbein field we can regard the equations of states for an equilibrium thermodynamic system as
the generalized eikonal equations, which are equivalent to HJ equations.
Next, from the Hamiltonian \eqref{H} and Hamilton's equation of motion, we have
\begin{subequations}
\begin{empheq}[left=\empheqlbrace]{align}
   \frac{d q^{\mu}}{d \tau} &= \frac{\partial H}{\partial p_{\mu}}
   = \frac{ g^{\mu \nu} p_{\mu}}{E},
   \label{qdot} \\
  \frac{d p_{\mu}}{d \tau} &= -\frac{\partial H}{\partial q^{\mu}}
   = -\frac{1}{2 E} \, \frac{\partial g^{\nu \rho}}{\partial q^{\mu}} \, p_{\nu} p_{\rho}.
\end{empheq}
\end{subequations}
Substituting Eq. \eqref{qdot} into the transformed Hamiltonian \eqref{K}, we obtain the relation
\begin{align}
   \sqrt{ g_{\mu \nu} \frac{d q^{\mu}}{d \tau} \frac{d q^{\nu}}{d \tau} } = 1,
\end{align}
which implies that
\begin{align}
   d \tau^2 =  g_{\mu \nu} dq^{\mu} dq^{\nu} .
  \label{dtau2}
\end{align}
Consequently, the "time" parameter $\tau$ in this setting gives a natural distance (arc-length) between equilibrium states of the thermodynamic systems on the manifold $\mathcal{M}$ equipped with the metric $g_{\mu \nu}$.

For the sake of simplicity, we only consider thermodynamical gas model
with $N=2$ dimensions \cite{Vaz}. A generalization for $N>2$ case is straightforward.
A thermal equilibrium system characterized by the specific entropy $s$
as a state function of the internal energy $u$ and volume $v$ per a molecule
of the gas.
We use the so-called \textit{entropy representation} $s = s(u, v)$, and
the first law of thermodynamics is expressed as
\begin{align}
  ds = \frac{1}{T} \, du + \frac{P}{T} \, dv,
\end{align}
where $T$ is the temperature and $P$ the pressure of the gas, respectively.
It is well known that they are related with
\begin{align}
  \frac{1}{T} = \left( \frac{\partial s}{\partial u} \right)_{v},\quad
  \frac{P}{T} = \left( \frac{\partial s}{\partial v} \right)_{u}.
\end{align}
Mathematically the above physical explanation is equivalent that the Pfaff equation
\begin{align}
 ds(u,v) -\frac{1}{T} \, du - \frac{P}{T} \, dv =0,
\end{align} 
is an exact differential, which was originally shown  by Carath\'eodory.

Next, introducing the Planck potential $\Xi$ given by
\begin{align}
  \Xi\left( \frac{1}{T}, \frac{P}{T} \right)
  := s(u, v) - \frac{1}{T} \, u - \frac{P}{T} \, v,
\end{align}
which is the total Legendre transform of the entropy $s(u,v)$.
These thermodynamical potentials and their variables correspond to
the potential functions and $\theta$- and $\eta$-coordinates in IG, respectively as follows:
\begin{subequations}
\label{correspondence}
\begin{align}
 \textrm{the $\eta$-coordinates: } &\quad \eta_1 \corrto u, \; \eta_2 \corrto v, \\
 \textrm{the $\theta$-coordinates: }&\quad \theta^1 \corrto -\frac{1}{T}, \; \theta^2 \corrto -\frac{P}{T}, \\
 \textrm{the $\theta$-potential: } &\quad \Psi(\bm{\theta}) \corrto \Xi\left( \frac{1}{T}, \frac{P}{T} \right), \\
\textrm{the $\eta$-potential: } &\quad \Psi^{\star}(\bm{\eta}) \corrto -s(u, v).
\label{eta-potential}
\end{align}
\end{subequations}
The two potentials are of course related by the total Legendre transformation 
\begin{align}
  \Psi(\theta^1, \theta^2) = \theta^{\mu} \eta_{\mu} - \Psi^{\star}(\eta_1, \eta_2),
\end{align}
and
\begin{align}
  \eta_{\mu} = \frac{\partial}{\partial \theta^{\mu}} \, \Psi(\bm{\theta}), \quad 
\theta^{\mu} = \frac{\partial}{\partial \eta_{\mu}} \, \Psi^{\star}(\bm{\eta}), \quad \mu=1,2.
\label{theta}
\end{align}

An example of Maxwell's relations in thermodynamics is described by 
\begin{align}
  \frac{\partial}{\partial v} \, \left(\frac{1}{T} \right)
   = \frac{\partial}{\partial u} \, \left( \frac{P}{T} \right),
\end{align}
which is equivalent to the integrability condition
$\partial \theta^1 / \partial \eta_2 = \partial \theta^2 / \partial \eta_1$. By using the relation \eqref{theta}, this condition leads to  the
following relation for the $\eta$-potensial \eqref{eta-potential},
\begin{align}
  \frac{\partial^2}{\partial v \partial u} \,  \Big( -s(u,v) \Big) 
  = \frac{\partial^2}{\partial u \partial v} \, \Big( -s(u,v) \Big),
\end{align}
i.e., the Hessian matrix of the negentropy $-s(u,v)$ must be symmetric.
It is known \cite{Amari} that a metric obtained by the Hessian of a convex potential leads 
to the dually flat structure in IG.  Such a Hessian metric in thermodynamics is proposed by Ruppeiner \cite{Ruppeiner}. The Ruppeiner metric is written by
\begin{align}
   (g^{\rm R})_{\mu \nu} =  \frac{\partial^2}{\partial q^{\mu} \partial q^{\nu}} \, \big( -s \, \big),
\quad \mu,\nu=u, v,
\end{align}
where $q^u = u$ and $q^v = v$.

\subsection{Ideal gas model}
Ideal gas model is the simplest thermodynamic model of an dilute gas, which consists of
a huge number of molecule. It is described by
\begin{align}
   u = \frac{f}{2} \, k_{\rm B} T, \quad  P \, v = k_{\rm B} T,
 \label{IdGas}
\end{align}
where $u$ stands for the specific internal energy, $v$ the specific volume, $ k_{\rm B}$ Boltzmann constant,
$T$ the temperature, $P$ the pressure of the ideal gas. The parameter $f$ stands for the degree of freedom of a molecule of the gas, e.g., for a monatomic molecule $f=3$. The first equation of \eqref{IdGas} is the equipartition theorem and the second is the equation of state of the ideal gas model. It is known that the corresponding entropy $s(u,v)$ can be expressed as
\begin{align}
   s(u, v) = P_u \ln \frac{u}{u_0} + P_v \ln \frac{v}{v_0}.
  \label{s}
\end{align}
Here we introduced the reference state $(u_0, v_0)$ which satisfy $s(u_0, v_0)=0$,
and 
\begin{align}
  P_u := \frac{f}{2} \, k_{\rm B}, \quad P_v := k_{\rm B}.
\end{align}
Note that $P_u$ and $P_v$ are constants (conserved quantities) and canonical transformed momenta in \eqref{Q_P}.

The equations \eqref{IdGas} can be cast into the form \eqref{vielbein}, i.e.,
\begin{align}
\begin{pmatrix}
   u \; & 0 \\[1ex]
   0 & v
\end{pmatrix}
\begin{pmatrix}
   \frac{1}{T} \\[1ex]
   \frac{P}{T} 
\end{pmatrix}
=
\begin{pmatrix}
\frac{f}{2} \, k_{\rm B}\\[1ex]
 k_{\rm B}
\end{pmatrix},
\end{align}
with
\begin{align}
e_i{}^{\mu} =
\begin{pmatrix}
   u & 0 \\
   0 & v
\end{pmatrix},
\label{2veinIG}
\quad p_{\mu} =
\begin{pmatrix}
   \frac{1}{T} \\[1ex]
   \frac{P}{T} 
\end{pmatrix}
, \quad
r_i =
\begin{pmatrix}
P_u\\[1ex]
P_v
\end{pmatrix},
\end{align}
Let us introduce the diagonal metric tensor $\eta$ as
\begin{align}
\eta^{i j} =
\begin{pmatrix}
   \frac{1}{\alpha^2} & 0 \\
   0 & \frac{1}{\beta^2}
\end{pmatrix}, \quad
\eta_{i j} =
\begin{pmatrix}
   \alpha^2 & 0 \\
   0 & \beta^2
\end{pmatrix},
\end{align}
where $\alpha^2$ and $\beta^2$ are scale factors \cite{scaleF}.
Then, from Eq. \eqref{R-metric}, the corresponding inverse matrix of the metric matrix becomes 
\begin{align}
   g^{\mu \nu} = \begin{pmatrix}
   \frac{u^2}{\alpha^2} & 0 \\[1ex]
   0 & \frac{v^2}{\beta^2}
\end{pmatrix}.
  \label{gIdGas}
\end{align}
Next, from the relations \eqref{p} and \eqref{R-metric}, we have
\begin{align}
  H &= \sqrt{g^{\mu \nu} p_{\mu} p_{\nu}} = \sqrt{\eta^{i j}  e_i{}^{\mu} p_{\mu}  e_j{}^{\nu} p_{\nu}}
  = \sqrt{\eta^{i j}  r_i  r_j} = E,
\end{align}
then we find the explicit expression of constant $E$ as
\begin{align}
   E(P_u, P_v) = \sqrt{ \frac{(P_u)^2}{\alpha^2} + \frac{(P_v)^2}{\beta^2} }.
\end{align}
The corresponding generalized eikonal equation \eqref{GEikonal} becomes
\begin{align}
 \frac{u^2}{\alpha^2} \left(\frac{\partial W}{\partial u} \right)^2 +   \frac{v^2}{\beta^2} \left(\frac{\partial W}{\partial v} \right)^2  = \frac{(P_u)^2}{\alpha^2} + \frac{(P_v)^2}{\beta^2},
\end{align}
from which a complete solution $W$ can be obtained as
\begin{align}
  W(u, v, P_u, P_v) = P_u \ln u + P_v \ln v,
\end{align}
where the constant of integration is set to zero.
Consequently the action $S$ of the ideal gas model becomes
\begin{align}
   S(u,v; P_u, P_v, \tau) = P_u \ln u  + P_v \ln v - E(P_u, P_v) \tau,
  \label{S_idGas}
\end{align}
and the corresponding generating function \eqref{G} is
\begin{align}
   G(u,v, \tau; u_0, v_0, \tau_0) =  P_u \ln \frac{u}{u_0}  + P_v \ln \frac{v}{v_0} - E(P_u, P_v) (\tau-\tau_0),
\end{align}
which satisfies the relations
\begin{align}
   \frac{\partial G}{\partial P_i} = 0, \quad i=u, v,
\end{align}
since $\partial S(\bm{q}, \bm{P}, \tau) / \partial P_i = \partial S(\bm{q}_0, \bm{P}, \tau_0) / \partial P_i = Q^i$. Using the expression \eqref{S_idGas} of the action $S$ of the ideal gas model, it follows that
\begin{subequations}
\begin{empheq}[left=\empheqlbrace]{align}
  \frac{\partial G}{\partial P_u} &= \ln \frac{u}{u_0}  - \frac{P_u (\tau-\tau_0)}{\alpha^2 E}  =0,\\
\frac{\partial G}{\partial P_v} &= \ln \frac{v}{v_0}  - \frac{P_v (\tau - \tau_0) }{\beta^2 E} =0,
\end{empheq}
\end{subequations}
By choosing $\alpha^2=P_u$ and $\beta^2=P_v$, which corresponds to a constant pressure process
as explained in \ref{B},
we have
\begin{align}
  u(\tau) &= u_0 \exp \left[ \frac{(\tau-\tau_0)}{E} \right], \quad
  v(\tau) = v_0 \exp \left[ \frac{(\tau - \tau_0)}{E} \right].
\end{align}
and
\begin{align}
  \frac{d u}{d\tau} &= \frac{u}{E}, \quad
  \frac{d v}{d\tau} = \frac{v}{E},
 \label{GFlowIdGas}
\end{align}

Next let us consider the gradient flow of $u(t)$ and $v(t)$.
From the correspondence relations \eqref{correspondence}, the direct mapping of the gradient flow equations \eqref{grad-eqs} in IG to those in the ideal gas model become
\begin{align}
  \frac{d u}{dt} &= g^{uu} \, \frac{\partial s(u,v)}{\partial u}, \quad
 \frac{d v}{dt} = g^{vv} \, \frac{\partial s(u,v)}{\partial v}.
\end{align}
Substituting the expression \eqref{gIdGas} and entropy \eqref{s} into these equations,
we find 
\begin{align}
  \frac{d u}{d t} = \frac{P_u}{\alpha^2} \, u = u, \quad
 \frac{d v}{d t} = \frac{P_v}{\beta^2} \, v = v.
\label{GradFlowIG}
\end{align}
for a constant pressure $P$ process.
Comparing the results \eqref{GFlowIdGas} and \eqref{GradFlowIG} we find 
\begin{align}
    d \tau = E \, d t. 
 \label{dtau}
\end{align}

Next,
taking the derivative of the specific entropy \eqref{s}, we have
\begin{align}
 ds(u,v) = \frac{1}{T} \, du + \frac{P}{T} \, dv,
\label{ds}
\end{align}
which corresponds to the relation
\begin{align}
   d W(\bm{q}, \bm{P}) = \frac{\partial W}{\partial q^i} \, dq^i +  \frac{\partial W}{\partial P_i} \, dP_i
  = p_i dq^i, 
\end{align}
in analytical mechanics, because $p_i = \partial W / \partial q^i$ in \eqref{p_Q} and $dP_i = 0$ in \eqref{Q_P}. The relation \eqref{dW} in analytical mechanics corresponds to
\begin{align}
  ds(u, v)  = E \, d\tau,
  \label{ds_e_tau}
\end{align}
in thermodynamics.

From the equations \eqref{IdGas}, we have
\begin{align}
  \frac{1}{T} du = - u \, d \left(\frac{1}{T} \right), \quad
  \frac{P}{T} dv = - v \, d \left(\frac{P}{T} \right).
\end{align}
Substituting these relations into Eq. \eqref{ds}, it follows that
\begin{align}
 ds(u,v) =  - u \, d \left(\frac{1}{T} \right) - v \, d \left(\frac{P}{T} \right).
\end{align}
For a constant pressure process, this becomes
\begin{align}
 ds(u,v) = (u + v P) \, \frac{dT}{T^2}.
\end{align}
Since $u = P_u T, v P = P_v T$ and $E = \sqrt{P_u + P_v}$, we have
\begin{align}
 ds(u,v) =   (P_u + P_v) \, \frac{dT}{T} = E^2 \, d \ln T.
\end{align}
Comparing this relation, \eqref{dtau} and \eqref{ds_e_tau}, we finally obtain that
\begin{align}
   d\tau = E \, d\ln T, \quad dt = d\ln T.
\label{dtau_dt}
\end{align}
In this way, we find out the relation between the parameter $t$ in the gradient flows in IG and
the temperature $T$ in a simple gas model of thermodynamics. 

\subsection{Van der Waals gas}
\label{sec:VWgas}

Since the ideal gas model is simple, we next consider the more realistic gas model by van der Waals.
His famous model
 is characterized by the following equations of states.
\begin{align}
  u + \frac{a}{v} = \frac{f}{2} \, k_{\rm B} T, \quad  (P + \frac{a}{v^2})(v-b) = k_{\rm B} T.
 \label{VdWGas}
\end{align}
The first equation of \eqref{VdWGas} states the equipartition theorem and the second equation states the equation of state by van der Waals \cite{van_der_Waals}. 
The term $a/v^2$ accounts for long-range attractive forces which increase pressure, and the $b$ term accounts for short range repulsive forces which decrease the volume available to molecules. In the limit of $a =b=0$, the gas model of van der Waals reduces to the ideal gas model. 

Now, the equations in \eqref{VdWGas} can be cast into the form \eqref{vielbein} with
\begin{align}
(e^{\rm vw})_i{}^{\mu} = \begin{pmatrix}
   u + \frac{a}{v} & 0 \\[1ex]
   \frac{a}{v^2} (v-b) & \; \; v-b
\end{pmatrix},
\label{e_vW}
\quad p_{\mu} =
\begin{pmatrix}
   \frac{1}{T} \\[1ex]
   \frac{P}{T} 
\end{pmatrix},
\end{align}
and the constants
\begin{align}
r_i =
\begin{pmatrix}
\frac{f}{2} k_{\rm B}\\[1ex]
 k_{\rm B}
\end{pmatrix}
=
\begin{pmatrix}
P_u\\[1ex]
P_v
\end{pmatrix},
\end{align}
where we choose the conserved quantity $P_u = f k_{\rm B} / 2$ and $P_v = k_{\rm B}$.

The corresponding entropy $s^{\rm vw}(u,v)$ can be expressed as
\begin{align}
   s^{\rm vw}(u, v) = \frac{f}{2} k_{\rm B}\ln \left(\frac{u+\frac{a}{v}}{u_0 + \frac{a}{v_0}} \right) 
   + k_{\rm B} \ln \left( \frac{v-b}{v_0-b} \right),
  \label{s_vw}
\end{align}
where we set $s(u_0, v_0)=0$.
From Eq. \eqref{R-metric} the corresponding metric becomes 
\begin{align}
   (g^{\rm vw} )^{\mu \nu} = \begin{pmatrix}
   \frac{1}{\alpha^2} \left(u + \frac{a}{v} \right)^2 +  \frac{a^2 (v-b)^2}{\beta^2 v^4}  &  \quad \frac{a (v-b)^2}{\beta^2 v^2}  \\[1ex]
   \frac{a (v-b)^2}{\beta^2 v^2}  & \frac{(v-b)^2}{\beta^2} 
\end{pmatrix}.
  \label{gVdWGas}
\end{align}
Then the generalized eikonal equation becomes
\begin{align}
 \frac{\left( u + \frac{a}{v}  \right)^2}{\alpha^2}  \left(  \frac{\partial W}{\partial u}  \right)^2
 +   \frac{(v-b)^2}{\beta^2}  \left( \frac{a}{v^2} \frac{\partial W}{\partial u}
 + \frac{\partial W}{\partial v}  \right)^2 = (E^{\rm vw})^2.
 \label{vWeikonal}
\end{align}

Recall that the action $S$ is expressed in the form \eqref{action} and
we can choose the scale factors $\alpha^2$ and $\beta^2$ arbitrarily.
By solving generalized eikonal equation \eqref{vWeikonal} we can obtain a complete solution $W(u, v, P_u, P_v)$, and then the action $S$ is calculated as
\begin{align}
   S(\bm{q}, \bm{P}, \tau) = P_u \ln \left(u + \frac{a}{v} \right)  + P_v \ln (v-b)  - E^{\rm vw}(P_u, P_v) \; \tau.
\end{align}
Then, we obtain the following relations.
\begin{subequations}
\begin{empheq}[left=\empheqlbrace]{align}
  \frac{\partial G}{\partial P_u} &= \ln \frac{u+\frac{a}{v}}{u_0+\frac{a}{v_0}}  
  - \frac{P_u (\tau-\tau_0)}{\alpha^2 E^{\rm vw}}  =0,\\
\frac{\partial G}{\partial P_v} &= \ln \frac{v-b}{v_0-b}  - \frac{P_v (\tau - \tau_0) }{\beta^2 E^{\rm vw}} =0,
\end{empheq}
\end{subequations}
Hence we obtain the "time" $\tau$-dependence of $u$ and that of $v$ as
\begin{subequations}
\begin{empheq}[left=\empheqlbrace]{align}
  u(\tau) &= -\frac{a}{v(\tau)} + \left(u_0 + \frac{a}{v_0} \right) \exp \left[ P_u \frac{(\tau-\tau_0)}{\alpha^2 E^{\rm vw}} \right],\\
  v(\tau) &= b + (v_0 - b) \exp \left[ \frac{P_v (\tau - \tau_0)}{\beta^2 E^{\rm vw}} \right],
\end{empheq}
\end{subequations}
respectively. From these relations we find 
\begin{subequations}
\begin{empheq}[left=\empheqlbrace]{align}
  \frac{d u}{d \tau} &= \frac{P_u}{\alpha^2 E^{\rm vw}} \,\left(u + \frac{a}{v} \right) + \frac{P_v}{\beta^2 E^{\rm vw}} \frac{a}{v^2} (v-b), \\
 \frac{d v}{d \tau} &= \frac{P_v}{\beta^2 E^{\rm vw}} \, (v-b).
\end{empheq}
\label{GFE4vW}
\end{subequations}

Next let us consider the gradient flow of $u(t)$ and $v(t)$.
From the correspondence relations \eqref{correspondence}, the direct mapping of the gradient flow equations \eqref{grad-eqs} in IG to those in the gas model of van der Waals become
\begin{subequations}
\begin{empheq}[left=\empheqlbrace]{align}
  \frac{d u}{dt} &= (g^{\rm vw})^{uu} \, \frac{\partial s^{\rm vw}}{\partial u} 
        +(g^{\rm vw})^{uv} \, \frac{\partial s^{\rm vw}}{\partial v}, \\
 \frac{d v}{dt} &= (g^{\rm vw})^{vu} \, \frac{\partial s^{\rm vw}}{\partial u} 
        +(g^{\rm vw})^{vv} \, \frac{\partial s^{\rm vw}}{\partial v}.
\end{empheq}
\end{subequations}
Substituting the metric \eqref{gVdWGas} and entropy \eqref{s_vw} into these equations,
we find 
\begin{subequations}
\begin{empheq}[left=\empheqlbrace]{align}
  \frac{d u}{d t} &= \frac{P_u}{\alpha^2} \,\left(u + \frac{a}{v} \right) + \frac{P_v}{\beta^2} \frac{a}{v^2} (v-b), \\
 \frac{d v}{d t} &= \frac{P_v}{\beta^2} \, (v-b).
\end{empheq}
\label{GFEinIG}
\end{subequations}
Comparing the results \eqref{GFE4vW} and \eqref{GFEinIG} we find 
\begin{align}
    d \tau = E^{\rm vw} \, d t. 
\end{align}

It is known that the following canonical transformation \cite{BFM16} from the variables
$(\bm{q}, \bm{p})$ to $(\tilde{\bm{q}}, \tilde{\bm{p}})$ relates the ideal gas model and the gas model by van der Waals.
\begin{subequations}
\begin{align}
  q^u = u \to \tilde{q}^u = \tilde{u} := u + \frac{a}{v}, \quad 
  q^v = v \to \tilde{q}^v = \tilde{v} := v - b, \\
  p^u = \frac{1}{T} \to \tilde{p}^u = \frac{1}{\tilde{T}} := \frac{1}{T}, \quad 
  p^v = \frac{P}{T} \to \tilde{p}^v = \frac{\tilde{P}}{\tilde{T}} := \frac{P}{T} + \frac{a}{v^2} \frac{1}{T}.
\end{align}
\label{MathieuT}
\end{subequations}
This is so-called Mathieu transformation, which is a subgroup of canonical transformations
preserving the differential from
\begin{align}
  p_i \, d q^i = \tilde{p}_i \, d \tilde{q}^i.
\end{align}
Indeed we confirm that
\begin{align}
  \tilde{p}_u \, d \tilde{q}^u + \tilde{p}_v \, d \tilde{q}^v 
  &= \frac{1}{T} \, d\left( u + \frac{a}{v} \right) + \left(\frac{P}{T} + \frac{a}{v^2} \frac{1}{T} \right) \, d(v-b) \notag \\
 &= \frac{1}{T} \, du + \frac{P}{T} \, dv = p_u \, d q^u + p_v \, d q^v.
\end{align}
By applying the canonical transformation \eqref{MathieuT}, the equations \eqref{VdWGas} 
of the gas model by van der Waals become
\begin{align}
  \tilde{u} = \frac{f}{2} \, k_{\rm B} \tilde{T}, \quad  
  \tilde{P} \, \tilde{v} = k_{\rm B} \tilde{T}.
\end{align}
Consequently the relation \eqref{e_vW} is transformed as
\begin{align}
  (e^{\rm vw})_i{}^{\mu} \to  
\begin{pmatrix}
   \tilde{u} & 0 \\
    0 & \tilde{v}
\end{pmatrix},
\end{align}
which is equivalent to the first relation in \eqref{2veinIG} of the ideal gas model.

\begin{subequations}
\label{dudv}
\begin{align}
 \frac{d}{d \tau} \, \tilde{u} 
  &= \frac{P_u}{\alpha^2 E^{\rm vw}} \, \tilde{u}, \quad
\frac{d}{d \tau} \, \tilde{v}
  = \frac{P_v}{\beta^2 E^{\rm vw}} \, \tilde{v},
 \\
 \frac{d}{d \tau} \, \frac{\tilde{P}}{\tilde{T}} 
  &= - \frac{P_v}{\beta^2 E^{\rm vw}} \left( \frac{\tilde{P}}{\tilde{T}} \right), \quad
\frac{d}{d \tau} \,  \left( \frac{1}{\tilde{T}} \right) 
  = - \frac{P_u}{\alpha^2 E^{\rm vw}} \, \left( \frac{1}{\tilde{T}} \right).
\end{align}
\end{subequations}

As explained in \ref{B}, a constant effective pressure $\tilde{P} = P + a / v^2$ process
is achieved when we choose the scale parameter as $\alpha^2 = P_u$ and $\beta^2=P_v$. In this case,
the relations \eqref{dudv} become
\begin{subequations}
\begin{align}
 \frac{d}{d \tau} \, \tilde{u} 
  &= \frac{\tilde{u}}{E^{\rm vw}}, \quad
\frac{d}{d \tau} \, \tilde{v}
  = \frac{\tilde{v}}{E^{\rm vw}}, \\
   \frac{d}{d \tau} \, \frac{\tilde{P}}{\tilde{T}} 
  &= - \frac{1}{E^{\rm vw}} \left( \frac{\tilde{P}}{\tilde{T}} \right), \quad
\frac{d}{d \tau} \,  \left( \frac{1}{\tilde{T}} \right) 
  = - \frac{1}{E^{\rm vw}} \, \left( \frac{1}{\tilde{T}} \right).
\end{align}
respectively. Then we see that
\begin{align}
 ds^{\rm vw} &= \frac{1}{\tilde{T}} \, d \tilde{u} + \frac{\tilde{P}}{\tilde{T}} \, d \tilde{v}
  = \left( \frac{\tilde{u}}{\tilde{T}}  + \frac{\tilde{v} \tilde{P}}{\tilde{T}} \right) \frac{d \tau}{E^{\rm vw}}
  = \left( P_u + P_v \right) \frac{d \tau}{E^{\rm vw}} = E^{\rm vw} d \tau,
\end{align}
\end{subequations}
where in the last step we used $E^{\rm vw} = \sqrt{P_u + P_v}$.

\section{Discussion}
Having found out the nontrivial relation \eqref{dtau_dt} between the time parameter $t$ of the gradient flows in IG and the temperature $T$ of a simple gas model in thermodynamics, let us turn our focus on the gradient flow equation \eqref{GradflowEq}.

\subsection{On the double exponential $t$-dependency}
\label{dexpt}
First we'd like to answer the question concerning the double exponential $t$-dependency
of the solution $q(t)$ in \eqref{q(t)}, i.e.,  the physical meaning of the parameter $t$ in this solution $q(t)$ from the view point of our results obtained until the previous sections.

Let us begin with the well known canonical probability of a thermally equilibrium system,
\begin{align}
   p_i(\beta) = \frac{1}{Z(\beta)} \exp \left( -\beta \mathcal{E}_i \right), \quad i=1,2, \ldots, N,
  \label{pi}
\end{align}
where $\beta := 1/ (k_{\rm B} T)$ is \textit{coldness} or the inverse temperature, $Z(\beta)$ is the partition function, $\mathcal{E}_i$ is the energy level of $i$-th state.
We assume that each discrete energy level is ordered as $\mathcal{E}_0 < \mathcal{E}_1 < \ldots < \mathcal{E}_N$. Consequently in the high temperature limit ($\beta \to 0$), every probability $p_i$ becomes equal, i.e.,
\begin{align}
  \forall i. \quad p_i (0) = \frac{1}{N}.
\end{align}
It is well known that the average energy $U$ is related to the partition function $Z(\beta)$ as
\begin{align}
   U := \sum_{i=1}^N  p_i(\beta) \mathcal{E}_i  = -\frac{d}{d \beta} \ln Z(\beta).
   \label{U}
\end{align}
Taking the logarithm of the both sides of \eqref{pi} we have
\begin{align}
   \ln p_i(\beta) = -\beta \mathcal{E}_i  - \ln Z(\beta).
  \label{ln_pi}
\end{align}
Taking the expectation of this with respect to $p_i(\beta)$ leads to
\begin{align}
   \sum_{i=1}^n p_i(\beta) \, \ln p_i(\beta) = -\beta U  - \ln Z(\beta).
  \label{potential}
\end{align}
Differentiating the both sides of \eqref{ln_pi} with respect to $\beta$ and using the relation \eqref{U},
if follows that
\begin{align}
 \frac{d}{d \beta}   \ln p_i(\beta) = -\mathcal{E}_i - \frac{d}{d \beta} \ln Z(\beta)
 = -\mathcal{E}_i +U,
  \label{relation}
\end{align}
Multiplying the both sides of \eqref{relation} by $\beta$, we have 
\begin{align}
 \beta \frac{d}{d \beta}  \ln p_i(\beta) = 
  - \beta \mathcal{E}_i +  \beta U,
  \label{b}
\end{align}
By subtracting \eqref{potential} from \eqref{ln_pi}, the right hand side of \eqref{b} is rewritten as
\begin{align}
 \beta \frac{d}{d \beta}  \ln p_i(\beta) &= \ln p_i(\beta) -\sum_{i=1}^N p_i(\beta) \, \ln p_i(\beta)
\end{align}
Since $p_0 := \lim_{\beta \to 0} p_i(\beta)$ is constant, we can rewrite this relation as
\begin{align}
 \frac{d}{d \ln \beta}  \ln \left(\frac{p_i(\beta)}{p_0} \right) &= 
  \ln \left( \frac{p_i(\beta)}{p_0} \right) -\sum_{i=1}^N p_i(\beta) \, \ln  \left( \frac{p_i(\beta)}{p_0} \right). 
\label{rel}
\end{align}
From the second relation in \eqref{dtau_dt} we see that
\begin{align}
  d \ln \beta = -d \ln \left( k_{\rm B} T \right) 
= - dt.
\end{align}
By integrating this relation, we can relate the time parameter $t$ and the coldness $\beta$ as
\begin{align}
   \beta = \exp( -t ) + C,
\end{align}
where $C$ is a constant of integration.
Setting $C = 0$ for simplicity, and substituting $\beta = \exp(-t)$ into \eqref{rel} we finally obtain that
\begin{align}
 \frac{d}{d t}  \ln \left(\frac{p_i({\rm e}^{-t})}{p_0} \right) &= - \left\{ 
  \ln \left( \frac{p_i({\rm e}^{-t})}{p_0} \right) -\sum_{i=1}^N p_i({\rm e}^{-t})) \, \ln  \left( \frac{p_i({\rm e}^{-t})}{p_0} \right) \right\}.
\end{align}
This is equivalent to the gradient flow equation in \eqref{GradflowEq} when we make the following
associations.
\begin{subequations}
\begin{align}
  \textrm{evolutional parameter:} \quad t &\Leftrightarrow -\ln \beta, \\
  \textrm{density: } q(t) &\Leftrightarrow  \textrm{probability: }  p_i(\beta), \\
  q_2: = \lim_{t \to \infty} q(t) &\Leftrightarrow p_0 := \lim_{\beta \to 0} p_i(\beta), \\
\mathrm{D}(q(t) \Vert q_2) := \Ev{q(t)}{ \ln \frac{q(t)}{q_2} } &\Leftrightarrow \sum_{i=1}^N p_i(\beta) \, \ln  \left( \frac{p_i(\beta)}{p_0} \right) = \mathrm{D}(p_i(\beta) \Vert p_0).
\end{align}
\end{subequations}
Therefore the double exponential $t$-dependency $\exp( \exp(-t))$ in the solution $q(t)$ in \eqref{q(t)} can be explained by these associations.  The evolutional parameter $t$ in \eqref{GradflowEq} is related to the coldness $\beta$ evolution through the relation $\beta = \exp(-t)$.

Next, let us turn our focus on 
Gompertz function $f_{\rm G}(t)$ defined by 
\begin{align}
  f_{\rm G}(t) := K \exp\big[ c \, \exp(-t) \big],
  \label{Gfunc}
\end{align}
where $c$ and $K$ are positive constants. This function $f_{\rm G}(t)$ is a sigmoid function, and has the double exponential $t$-dependency.
Gompertz \cite{Gompertz} studied human mortality for working out a series of tables mortality, and this suggested to him his law of human mortality in which he assumed that the mortality rate decreases exponentially as a person ages. Gompertz function is nowadays used in many areas to model a time evolution where growth is slowest at the start and end of a period.
The rule of his model is called \textit{Gompertz rule} which states that 
\begin{align}
   \frac{d}{dt} f_{\rm G}(t) = -f_{\rm G}(t) \ln \frac{f_{\rm G}(t)}{K},
   \label{Grule}
\end{align}
The solution of the Gompertz rule is \eqref{Gfunc}
if we set $K = \lim_{t \to \infty} f_{\rm G}(t)$ and $c = \ln (f_{\rm G}(0)  / K )$.

Now we observe that $q(t)$ in Eq. \eqref{q(t)} is proportional to Gompertz function. Indeed, let us define $Q(t)$ by the un-normalized version of $q(t)$,
\begin{align}
   Q(t) := q(t) \, \mathrm{e}^{\Psi(t)} = \exp\left[ {\mathrm{e}^{-t} \ln q_0 + (1-\mathrm{e}^{-t}) \ln q_2} \right]=
q_2 \, \mathrm{e}^{ \ln \left( \frac{q_0}{q_2} \right)  \, \mathrm{e}^{-t}}.
\end{align}
We see that this function $Q(t)$ is a Gompertz function \eqref{Gfunc} with $K=q_2$ and $c= \ln(q_0/q_2)$.
Then we have
\begin{align}
  \frac{d}{dt} \ln Q(t) = - \mathrm{e}^{-t} \ln \frac{q_0}{q_2} 
  = - \ln \frac{Q(t)}{q_2}, 
\end{align}
which is nothing but the Gompertz rule \eqref{Grule}
\begin{align}
  \frac{d}{dt} Q(t)  = - Q(t) \, \ln \frac{Q(t)}{q_2}. 
\end{align}

\subsection{The gradient flow and Hamilton flow}
\label{subsec}

Next we focus our attention on the relation between the gradient flow \eqref{grad-eqs} and Hamilton flow.
In this subsection, for the sake of clarification, we don't use Einstein's summation convention.

Let us consider the simple model described by the following characteristic function $W( \bm{q}, \bm{P})$,
\begin{align}
   W( \bm{q}, \bm{P}) =  \sum_{\mu=1}^m P_{\mu} \ln q^{\mu},
\label{Wa}
\end{align}
on a smooth manifold $\mathcal{M}$ with $m$-dimension.
Here each $P_{\mu}$ is a constant of motion.
The corresponding generalized momentum is given by
\begin{align}
  p_{\mu} = \frac{\partial W}{\partial q^{\mu}} = \frac{P_{\mu}}{q^{\mu}}, \quad \mu=1,2, \ldots, m.
\end{align}
Consequently  each constant of motion $P_{\mu}$ satisfies that
\begin{align}
  P_{\mu} = q^{\mu} \, p_{\mu}, \quad \mu=1,2, \ldots, m.
  \label{P}
\end{align}
The metric tensor is obtained by
\begin{align}
   g_{\mu \nu}(\bm{q}) = - \frac{ \partial^2 W}{\partial q^{\mu} \partial q^{\nu}} =  \frac{P_{\mu}}{(q^{\mu})^2}  \, \delta^{\mu \nu},  \quad \mu, \nu=1,2, \ldots, m,
\end{align}
and the inverse matrix of the metric is
\begin{align}
   g^{\mu \nu}(\bm{q}) = \frac{(q^{\mu})^2}{ P_{\mu}} \, \delta^{\mu \nu},  \quad \mu, \nu=1,2, \ldots, m,
\end{align}
Then the corresponding Hamiltonian \eqref{H} becomes
\begin{align}
  H(\bm{q}, \bm{p}) = \sqrt{ \sum_{\mu, \nu} g^{\mu \nu} (\bm{q}) \, p_{\mu} p_{\nu}} 
   = \sqrt{ \sum_{\mu} \frac{(q^{\mu} \, p_{\mu} )^2}{P_{\mu}}  },
\end{align}
and its value $E( \bm{P})$ is constant because this Hamiltonian has no explicit time dependence.
Indeed, by using the relations \eqref{P}, it follows that
\begin{align}
  E( \bm{P})  = \sqrt{ \sum_{\mu} P_{\mu} }.
\end{align}
Then Hamilton's equations of motion leads to
\begin{subequations}
\begin{empheq}[left=\empheqlbrace]{align}
  \frac{d}{d\tau} \, q^{\mu} &= \frac{\partial H}{\partial p_{\mu}} =  \frac{ (q^{\mu})^2 \, p_{\mu}}{ P_{\mu} E} = \frac{q^{\mu}}{ E}, \\
\frac{d}{d\tau} \, p_{\mu} &= -\frac{\partial H}{\partial q^{\mu}} 
       = -\frac{q^{\mu} \, (p_{\mu})^2}{ P_{\mu} E} = -\frac{p_{\mu}}{ E},
\end{empheq}
\label{EoM}
\end{subequations}
where we used the relations \eqref{P} again.
Now by using the relation \eqref{dtau},
the equations of motion \eqref{EoM} become
\begin{subequations}
\begin{empheq}[left=\empheqlbrace]{align}
  \frac{d}{d t} \, q^{\mu} &= q^{\mu}, \\
\frac{d}{d t} \, p_{\mu} &= -p_{\mu}.
\end{empheq}
\label{Hamflow}
\end{subequations}
We remind that the correspondence between IG and analytical mechanics:
\begin{subequations}
\begin{alignat}{2}
  \eta_{\mu} &\corrto q^{\mu}, &\qquad \theta^{\mu} &\corrto -p_{\mu}, \\
  \Psi^{\star}(\bm{\eta}) &\corrto  -W(\bm{q}, \bm{P}), &\quad
  \theta^{\mu} = \frac{\partial \Psi^{\star}(\bm{\eta})}{\partial \eta_{\mu}} &\corrto
  -p_{\mu} =  -\frac{\partial W(\bm{q}, \bm{P})}{\partial q^{\mu}}.
\end{alignat}
\end{subequations}
Then we see that the Hamilton flow described by \eqref{Hamflow} is equivalent to the gradient flow 
in IG described by the gradient equations \eqref{grad-eqs}.

Finally let us comment on the Lemma 3.3 in Ref. \cite{BN16} by Boumuki and Noda, where they introduced the potential functions for a dually flat space in IG as
\begin{align}
  \Psi^{\star}(\bm{\eta}) = - \sum_i \ln \eta_i,  \quad \eta_i = -1/ \theta^i, \quad i=1,2, \ldots, m.
  \label{BN}
\end{align}
This mathematical model is simple but non-trivial, since its gradient flow equation is reformulated as Hamilton equation. For the details see Ref. \cite{BN16}.

Now, from the above correspondence between IG and analytical mechanics, we note that
the first relation in \eqref{BN} is the special case in which all $P_{\mu}$ are set to unity in \eqref{Wa}, and the second relation
comes from the relation \eqref{P}.  As a result, their model \eqref{BN} is a special case of the model discussed in this subsection.



\section{Conclusions}

We have studied the gradient flows in IG  as the dynamics of a simple thermodynamic system.
The equations of states in thermodynamics are regarded as the generalized eikonal equations,
and we have incorporated a "time" ($\tau$) evolution  into thermodynamics as HJ dynamics. 
Through this "time" parameter $\tau$ in the HJ dynamics of the thermodynamics,  we have related 
the parameter $t$ in the gradient flow equation to the inverse temperature $\beta$ of the thermodynamic system
as shown in \eqref{dtau_dt}.
Based on this fact, we have found the physical origin of the double exponential $t$-dependency of the solution \eqref{q(t)} 
in subsection \ref{dexpt}. In this way, the gradient flow in IG is related to the HJ dynamics of the thermodynamical systems.

Thermodynamics is powerful and useful in a wide range of scientific fields.
Very recently, Ghosh and Bhamidipati \cite{GB19} studied the thermodynamics of black holes from the contact geometry point of view.
They showed that the thermodynamic processes of black holes can be modeled by characteristic curves of a suitable contact Hamilton-flow.
We hope our findings help further understandings not only between thermodynamics and IG but also among other different fields.

%


\section*{Acknowledgements}

The authors thank to Prof. G. Pistone for his interesting talk in
the SigmaPhi2017 Conference held in Corfu, Greece,
10-14 July 2017, and for organizing a seminar held at Collegio Carlo Alberto, Moncalieri, on 5 September 2017.
The first named author (T.W) also thanks to Dr. S. Goto for useful discussion on the early version of this work,
and to again Prof. G. Pistone for his kind invitation and discussions in Torino on 23rd January 2020.

The first named author (T.W.) is supported by Japan Society for the Promotion of Science (JSPS) Grants-in-Aid for Scientific Research (KAKENHI) Grant Number JP17K05341. The third named author (H. M.) is partially supported by the JSPS Grants-in-Aid for Scientific Research (KAKENHI) Grant Number JP15K04842 and JP19K03489.

\appendix


\section{Hamilton-Jacobi equation by canonical transformation}
\label{HJbyCT}
Here we briefly review the Carath\'{e}odory derivation \cite{Snow67} of HJ equation by canonical transformation. Recall that the non-uniqueness of Lagrangian, i.e.,  two Lagrangians which differ by a total derivative of some function $f(\bm{q}, t)$ with respect to time, describe the same system.
For example consider the following two Lagrangians $L$ and $L^*$ related through
\begin{align}
    \int_{t_a}^{t_b} \left( L(\bm{q}, \dot{\bm{q}}, t) + \frac{d f(\bm{q}, t)}{dt} \right) dt 
= \int_{t_a}^{t_b} \left( L^{\star}(\bm{Q}, \dot{\bm{Q}}, t) + \frac{d g(\bm{Q}, t)}{dt} \right) dt,
\end{align}
where $f(\bm{q}, t)$ and $g(\bm{Q}, t)$ are some functions.
Both Lagrangians lead to the same Euler-Lagrange equation, and consequently describe the same system.
Introducing the function $F(\bm{q}, \bm{Q}, t) := g(\bm{Q}, t) - f(\bm{q}, t)$, then
\begin{align}
    L(\bm{q}, \dot{\bm{q}}, t) = L^{\star}(\bm{Q}, \dot{\bm{Q}}, t) + \frac{d F(\bm{q}, \bm{Q}, t)}{dt}.
    \label{LandL*}
\end{align}

Now, introducing the Hamiltonians $H(\bm{q}, \bm{p}, t)$ and $K(\bm{Q}, \bm{P}, t)$ which
are Legendre duals of $ L(\bm{q}, \dot{\bm{q}}, t)$ and $ L^{\star}(\bm{Q}, \dot{\bm{Q}}, t) $, respectively, i.e.,
\begin{align}
  H(\bm{q}, \bm{p}, t) = p_{\mu} \frac{dq^{\mu}}{dt} - L(\bm{q}, \dot{\bm{q}}, t), \quad
  K(\bm{Q}, \bm{P}, t) = P_{\mu} \frac{dQ^{\mu}}{dt} - L^{\star}(\bm{Q}, \dot{\bm{Q}}, t).
\end{align}
By choosing $F(\bm{q}, \bm{Q}, t) =S(\bm{q}, \bm{P}, t) - P_{\mu}  \, Q^{\mu}$, then \eqref{LandL*} leads to
\begin{align}
    p_{\mu}  dq^{\mu} - H(\bm{q}, \bm{p}, t) dt  &= 
  P_{\mu}  dQ^{\mu} - K(\bm{Q}, \bm{P}, t) dt + d \big( S(\bm{q}, \bm{P}, t) - P_{\mu} Q^{\mu} \big) \notag \\
 &= -Q^{\mu} dP_{\mu} - K dt + \frac{\partial S}{\partial q^{\mu}} dq^{\mu} + \frac{\partial S}{\partial P_{\mu}} dP_{\mu} +
\frac{\partial S}{\partial t} dt,
\end{align}
which describes the canonical transformation from the canonical coordinates $(\bm{q}, \bm{p})$ to $(\bm{Q}, \bm{P})$. Since both the canonical coordinates are independent we obtain
\begin{align}
  p_{\mu} = \frac{\partial S}{\partial q^{\mu}}, \quad Q^{\mu} = \frac{\partial S}{\partial P_{\mu}}, \quad
  K = H + \frac{\partial S}{\partial t},
\end{align} 
and from Hamilton's equation of motion for the transformed Hamiltonian $K(\bm{Q}, \bm{P}, t)$
\begin{align}
   \frac{d}{dt} Q^{\mu} = \frac{\partial K}{\partial P_{\mu}}, \quad \frac{d}{dt} P_{\mu} = -\frac{\partial K}{\partial Q^{\mu}}.
\end{align}
We see that the transformed canonical coordinates  $(\bm{Q}, \bm{P})$ are conserved, i.e., $dQ^{\mu} / dt = dP_{\mu} /dt = 0$ when $K=0$. This condition leads to 
\begin{align}
  K = H( \bm{q}, \frac{\partial S}{\partial \bm{q}}, t) + \frac{\partial}{\partial t} \, S(\bm{q}, \bm{P}, t) = 0,
\end{align} 
which is HJ equation.

\section{Constant pressure process}
\label{B}

From Hamilton's equation of motion for \eqref{H}
with the metric \eqref{gIdGas}
of the ideal gas model, we obtain
\begin{subequations}
\begin{alignat}{2}
 \frac{d}{d \tau} u 
  &= \frac{P_u}{\alpha^2 E} u, &\qquad \qquad
\frac{d}{d \tau} v
  &= \frac{P_v}{\beta^2 E}  v, \label{1stEq} \\
 \frac{d}{d \tau} \left( \frac{P}{T}  \right)
  &= - \frac{P_v}{\beta^2 E} \left( \frac{P}{T}  \right), &\quad
\frac{d}{d \tau} \left( \frac{1}{T} \right)
  &= - \frac{P_u}{\alpha^2 E} \left( \frac{1}{T} \right).
\label{2ndEq}
\end{alignat}
\end{subequations}
From the first relation of  \eqref{2ndEq}, we have
\begin{align}
 \frac{d}{d \tau} P 
 = -T P \, \frac{d}{d \tau} \left( \frac{1}{T} \right)  - \frac{P_v P}{\beta^2 E}.
 \label{dPdtau}
\end{align}
Substituting the second relation of \eqref{2ndEq} into \eqref{dPdtau}, we obtain
\begin{align}
 \frac{d}{d \tau} P 
 = \frac{P}{E} 
       \left( \frac{P_u}{\alpha^2} - \frac{P_v}{\beta^2}  \right).
\end{align}
Thus when we choose the scaling factors as
\begin{align}
  \alpha^2 = P_u, \quad \beta^2 = P_v,
  \label{const-tildeP}
\end{align}
it is a constant pressure ($P$) process.

Next we consider the gas model by van der Waals.
From Hamilton's equation of motion for the Hamiltonian \eqref{H} with \eqref{gVdWGas}, we obtain
\begin{subequations}
\begin{align}
 \frac{d}{d t} \left( u + \frac{a}{v} \right)
  &= \frac{P_u}{\alpha^2 E} \left( u + \frac{a}{v} \right), \\
\frac{d}{d t} v
  &= \frac{P_v}{\beta^2 E} (v - b), \\
 \frac{d}{d t} \left( \frac{P}{T} + \frac{a}{v^2} \frac{1}{T} \right)
  &= - \frac{P_v}{\beta^2 E} \left( \frac{P}{T} + \frac{a}{v^2} \frac{1}{T} \right), \\
\frac{d}{d t} \left( \frac{1}{T} \right)
  &= - \frac{P_u}{\alpha^2 E} \left( \frac{1}{T} \right).
\end{align}
\end{subequations}
By using these equations, we obtain
\begin{align}
 \frac{d}{d \tau} \left( P + \frac{a}{v^2}  \right)
  &= \frac{1}{E} \left( P + \frac{a}{v^2}  \right) 
       \left( \frac{P_u}{\alpha^2} - \frac{P_v}{\beta^2}  \right).
\end{align}
Thus when we choose
\begin{align}
  \alpha^2 = P_u, \quad \beta^2 = P_v,
  \label{const-tildeP}
\end{align}
it corresponds to a constant $P + a / v^2$ process.
In the ideal gas limit, i.e., $a = b = 0$, it reduces to a constant pressure  (isobaric) process.

\section*{Conflicts of Interest}
The authors declare no conflict of interest.



\end{document}